\documentclass{iau} 
\usepackage{graphicx}

\title[Machine Learning in Large Spectral Archives]%%
{Identification of Interesting  Objects in Large Spectral Surveys Using Highly Parallelized Machine Learning}

\author[\v{S}koda \textit{et al.}]{Petr \v{S}koda$^1$, Andrej Pali\v{c}ka$^2$, Jakub Koza$^2$ \and Ksenia Shakurova$^2$}

\affiliation{$^1$Astronomical Institute of the Czech Academy of Sciences \\ Fri\v{c}ova 298, 251\,65 Ond\v{r}ejov, Czech Republic \\ email: {\tt skoda@sunstel.asu.cas.cz} \\[\affilskip]
$^2$Faculty of Information Technology, Czech Technical University in Prague \\ Th\'akurova 9, 160\,00 Prague 6, Czech Republic}

\pubyear{2017}
\volume{325}  %% insert here IAU Symposium No.
\jname{Astroinformatics 2016}
\editors{M. Brescia, S. Cavuoti, G.S. Djorgovski, \\ E. Feigelson \& G. Longo, eds.}
\begin{document}

\maketitle

\begin{abstract}
The current archives of LAMOST multi-object spectrograph contain millions of
fully reduced spectra, from which the automatic pipelines have produced
catalogues of many parameters of individual objects, including their
approximate spectral classification.  This is, however, mostly based on the
global shape of the whole spectrum and on integral properties of spectra in
given bandpasses, namely presence and equivalent width of prominent spectral
lines, while for identification of some interesting object types (e.g. Be stars
or quasars) the detailed shape of only a few lines is crucial. Here the machine
learning is bringing a new methodology capable of improving the reliability of
classification of such objects even in boundary cases.

We present results of Spark-based semi-supervised machine learning of LAMOST
spectra  attempting to automatically identify the single and double-peak
emission of H$_\alpha$ line typical for Be and B[e] stars.  The labelled sample
was obtained from archive of 2m Perek telescope at Ond\v{r}ejov observatory. A
simple physical model of spectrograph resolution was used in domain adaptation
to LAMOST training domain. The resulting list of candidates contains dozens of
Be stars (some are likely yet unknown), but  also a bunch of interesting
objects resembling spectra of quasars and even blazars, as well as many
instrumental artefacts.  The verification of a nature of interesting candidates
benefited considerably from cross-matching and visualisation in the Virtual
Observatory environment.  

\keywords{stars: emission-line, Be, surveys, methods:
statistical, techniques: spectroscopic}
\end{abstract}

\firstsection % if your document starts with a section,
\section{Introduction}

There are many objects in the Universe that may show interesting shapes of some
important spectral lines. The objects presenting emission lines, as  are Be
stars,  where a gaseous envelope in the shape of a sphere or a disk is expected
\cite{2003PASP..115.1153P}, and rare class of B[e] stars showing infrared
excess \cite{2003A&A...408..257Z}, are especially interesting, thanks to
their complicated physics.  The emission lines in spectra of such objects may
present under different physical conditions single peak, double peak with
different ratios of components or even complicated combined emission and absorption profiles
\cite{2010ApJS..187..228S}.  To find emission line objects with given shape of
spectral line in a big survey, the automatic procedure must be used based on
principles of supervised machine learning.
\section{Machine Learning}
Machine learning is the field of informatics, closely related to the advanced
statistical inference, which tries to build models of data  by learning from
sample inputs and make predictions based on such learned models.  It is divided
mainly into supervised and unsupervised methods with a number of subclasses. 
\begin{description}
\item[Supervised:]\  These algorithms have some prior knowledge about the data,
that was supplied by some external means. This is usually done by a human
domain expert. Main representatives are the \emph{classification} tasks, where the
models classify data into some predefined, finite set of classes, and \emph{regression}
tasks, which infer output of a real-valued function. 
\item[Unsupervised]\  Algorithms in this group do not have any prior knowledge
of the data. They attempt to discover corresponding relationships themselves. This is by
definition a much harder task than supervised learning, however it can potentially
lead to much more interesting results. Example tasks are \emph{clustering}, 
\emph{automated feature selection} or \emph{outliers detection}.
\end{description}
A special category is a \textbf{Semi-Supervised Learning}. It is a group of
algorithms designed to work on datasets, that have very few labelled data
points compared to the amount of unlabelled points.  It resembles a supervised
learning in that we have labelled data, however with an addition of also having
unlabelled data, which may help us estimate the distribution of the data set
more precisely.  There is, however, another possible variant of
semi-supervised learning. Here, we are not using labelled data, but merely some
constraints. These constraints may link some points, that share the same label,
or they may reveal the actual number of classes. This resembles unsupervised
learning, however with some a priori information about the data.

In big spectral archives, where it is almost impossible to investigate every
spectrum visually, the yet unknown rare objects with strange features, or even
sources with yet undiscovered physical mechanism may be in principle found
using machine learning. 
\section{ LAMOST Spectral Surveys}
The LAMOST telescope (\cite{Cui}) has been delivering  one of
currently largest mega-collections  of spectra (similar to Sloan Digital Sky
Survey). The sixteen LAMOST spectrographs are fed by 4000 fibres positioned by
micro-motors. Its publicly accessible data release DR1
(\cite{Luo}) contains altogether \textbf{2\,204\,696}
spectra, with a spectral resolving power $\sim 1800$, covering the range 
3690-9100\AA. The LAMOST pipeline classified \textbf{1\,944\,329} of them as
stellar ones.
\section{Ond\v{r}ejov CCD700  Archive}
\noindent The unique source of the spectra of emission line stars (mostly Be
and several B[e]) is the archive of spectra obtained with 700mm camera of the
coud\`e  spectrograph of the 2m Perek Telescope at Ond\v{r}ejov observatory, a
part of the Astronomical Institute of the Czech Academy of Sciences. The
archive (named CCD700) contains about twenty thousand spectra of mainly Be
stars and other emission-line objects exposed mostly in spectral range
6250--6700\AA{}  with spectral resolving power $\sim 13000$. 
\subsection{Cross-matching and Labelling of LAMOST and CCD700 Archives}

Using the technology of Virtual Observatory, namely the combination of Table Access
\cite{TAP} and Simple Spectra Access \cite{SSAP} protocols we have identified
less than ten objects (e.g. BT CMi or HD53416) observed both by Ond\v{r}ejov 2m
Perek Telescope and LAMOST. This limits the usage of normal supervised training
due to the lack of labelled spectra in LAMOST DR1 selected for training set. So
the semi-supervised method must be used.  In CCD700 archive we have visually
identified 1696 spectra with clearly defined classes of spectral shapes. The
most important part consisted of spectra of selected Be stars with different
spectral line profiles as double peak emission, emission with central
absorption or deep absorption with small single or double emission peak. All
such interesting cases were given label 1 (target class).  They were
complemented by absorption spectra of many classical stars labelled 0.  This
sample became a training set.  The spectra, however, could not be used
immediately due the the different spectral resolving power of CCD700 archive
and LAMOST.  We needed to perform a \textit{Domain Adaptation} of training
sample.

\subsection{Domain Adaptation of CCD700 Spectra}
\begin{figure}[!ht]
\begin{center}
 \includegraphics[width=\textwidth]{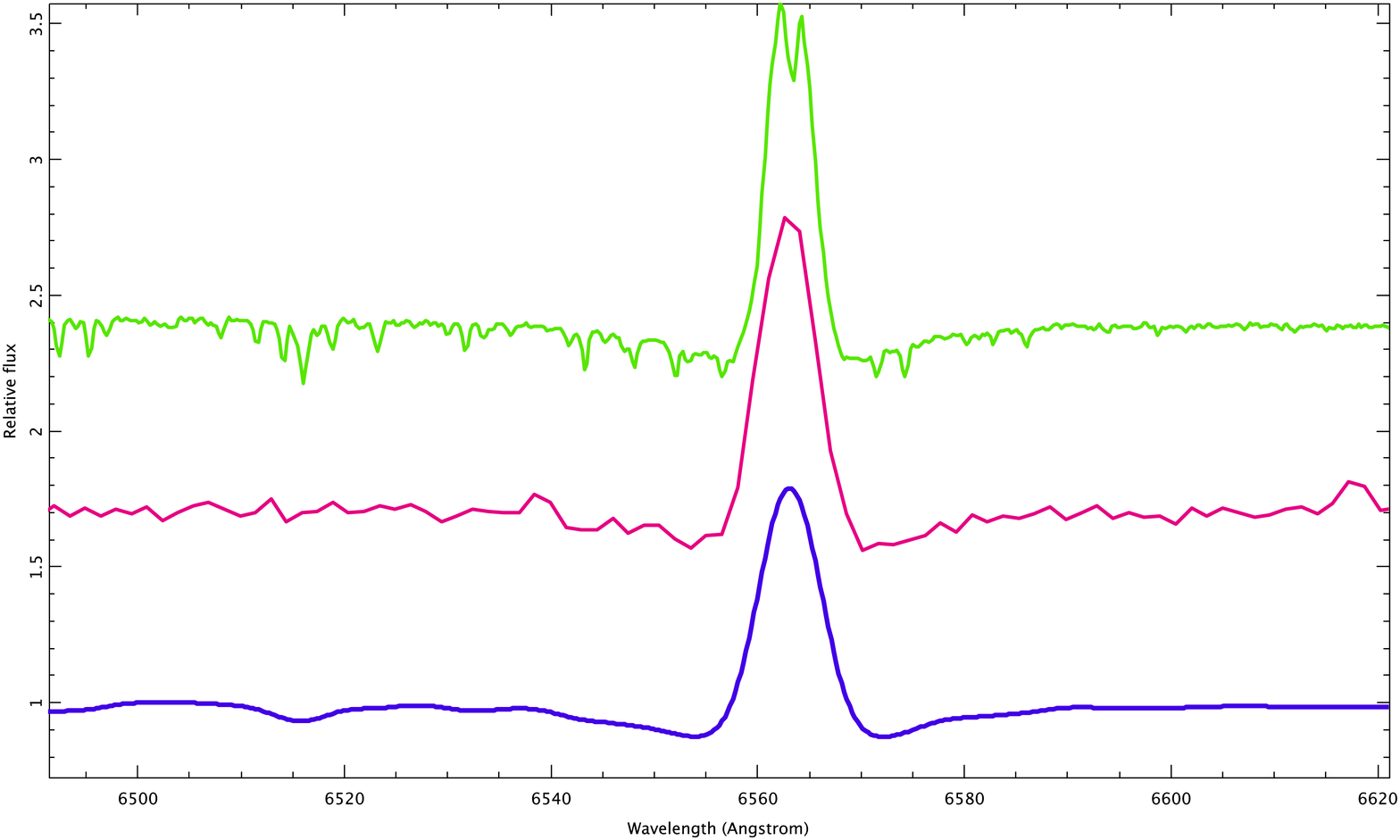} 
 \caption{Comparison of the original (top) and convolved (bottom) CCD700 spectrum of HD53416 with one observed by LAMOST (middle)}
   \label{comp_resolution}
\end{center}
\end{figure}

Before starting the semi-supervised learning, all Ond\v{r}ejov spectra with
higher resolution must be converted  to the same resolution as LAMOST data. In
other words the samples from one domain (CCD700) are transformed to second
domain (LAMOST), so that the CCD700 spectra of selected emission stars will
look like exposed with LAMOST spectrograph. Here we use the  model based on our
(simplified) physical knowledge of the principles how the spectrograph works.
The spectral resolving power degradation may be roughly approximated by the
convolution with the Gaussian kernel with Full Width at Half Maximum (FWHM)
proportional to the ratio of resolving powers. In our case we have used
Gaussian with FWHM of 5 pixels.  Finally, the spectrum has to be re-binned into
larger pixels, as the LAMOST spectrum contains less pixels over the same
spectral range. Such simulated spectra are treated as the original data for the
input preprocessing. See Fig.~\ref{comp_resolution} for an example of
resolution degradation. The top spectrum is an original CCD700 spectrum of
HD53416, the bottom one its convolved version (please notice the smearing of
the double peak profile), while the middle one is the same star observed with
LAMOST.
\section{Input Spectra Preprocessing}
An important part of data preparation before applying machine learning is the
data pre-processing. In our case all the spectra had to be normalised to the
continuum (rectified), cut to the same wavelength range 6250--6750\AA{} (for
CCD700 also convolved with Gaussian as explained above) and re-binned into the
same grid of wavelength points.  This gave us the number of so called Feature
Vectors (FV).  The result of the preprocessing is the  Comma-separate values
(CSV) file with all spectral intensities interpolated to the same wavelength
grid mapped to the individual elements of the FV. Here the value of physical
wavelength in \AA\ is unknown and must be reconstructed for visualisation from
the mapping metadata.  This CSV file is uploaded to a computing cluster running
the semi-supervised machine learning algorithms .
\section{Semi-Supervised Machine Learning }
We have used two graph-based methods, described in
\cite{chapelle2006semi}, which were adopted for parallelisation on Spark engine.
The full implementation details are explained in \cite{Palicka_MT}.
\begin{itemize}
\item[\textbf{Label Propagation}]
Label propagation is an algorithm that leverages the graph representation
of data to fit a model.  Labels are encoded as an one-hot variable, so that
we may support multi-class classification. The algorithm basically computes
weights for labels for each data point based on the distance to its
neighbours. Note that in our version of the algorithm, the initial
labels do not change and are reset to their original value in each step.
\item[\textbf{Label Spreading}]
is a similar algorithm to label propagation. It
uses the normalised graph Laplacian  to propagate the label
information across the graph. It also allows the labels to retain some partial
information from the initial labelling.
\end{itemize}
\section{Massively Parallelised Processing Using Spark}
The Apache Spark ({\tt http://spark.apache.org}) is a cluster computing
technology allowing the fast computation on number of computing nodes in
parallel. We have used the academical cluster MetaCentrum consisting of
twenty-four sixteen-core nodes (the number of nodes assigned by the system is
however unknown, dependent on a availability and load of the cluster). 
The data were distributed across all nodes by Hadoop Distributed File System ---
HDFS ({\tt http://hadoop.apache.org}). The search was run on more than
fifty thousand spectra randomly selected from those labelled as star by LAMOST DR1
pipeline.
\section{Results}
\begin{figure}[!ht]
\begin{center}
 \includegraphics[width=\textwidth]{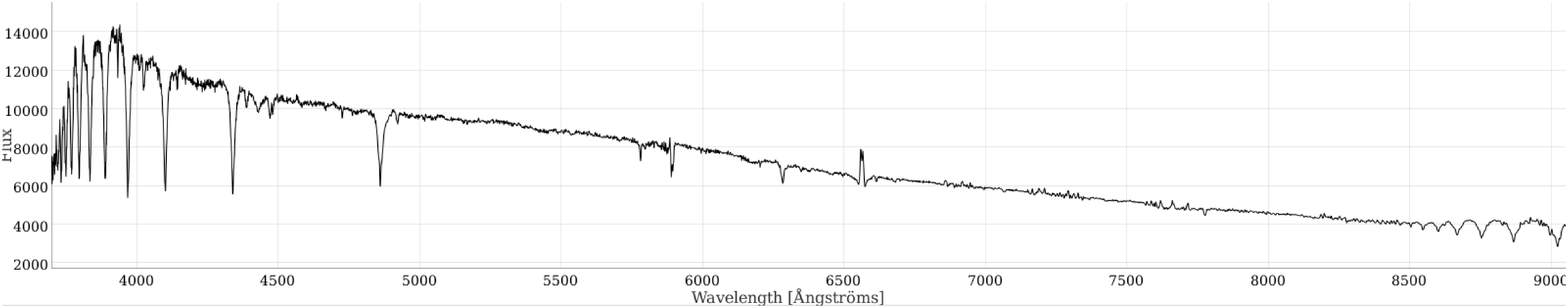} 
 \includegraphics[width=\textwidth]{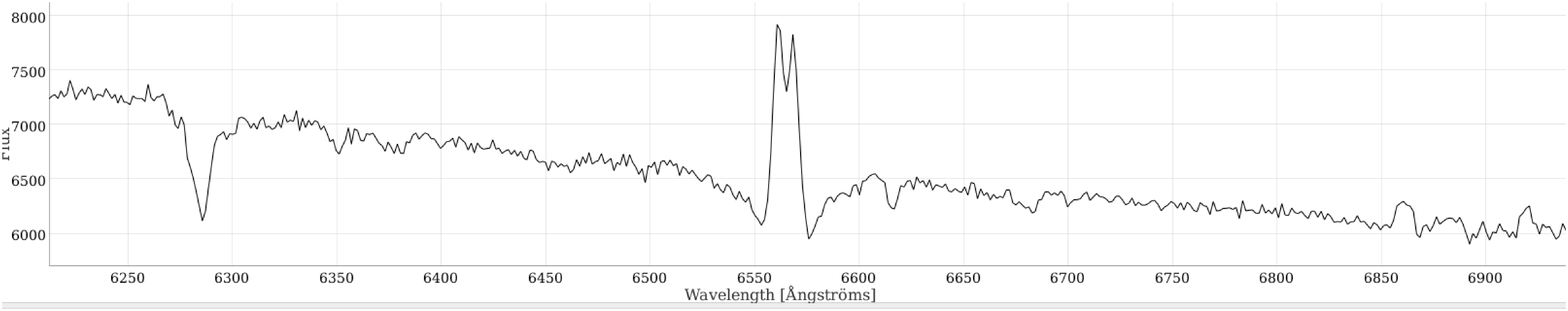} 
 \caption{Example of Be star identified in LAMOST DR1 by machine learning:\
 Strong double peak emission }
   \label{double_emission}
\end{center}
\end{figure}

\begin{figure}[!ht]
\begin{center}
 \includegraphics[width=\textwidth]{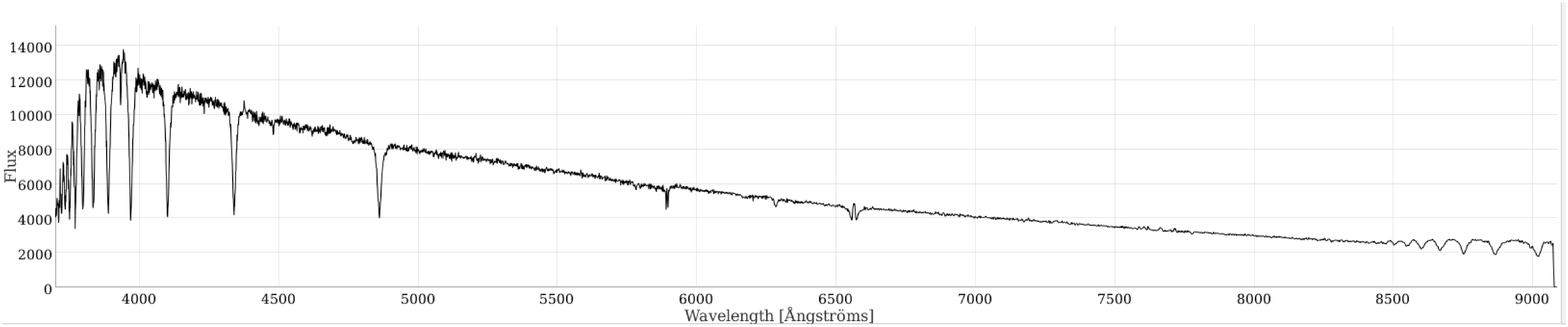} 
 \includegraphics[width=\textwidth]{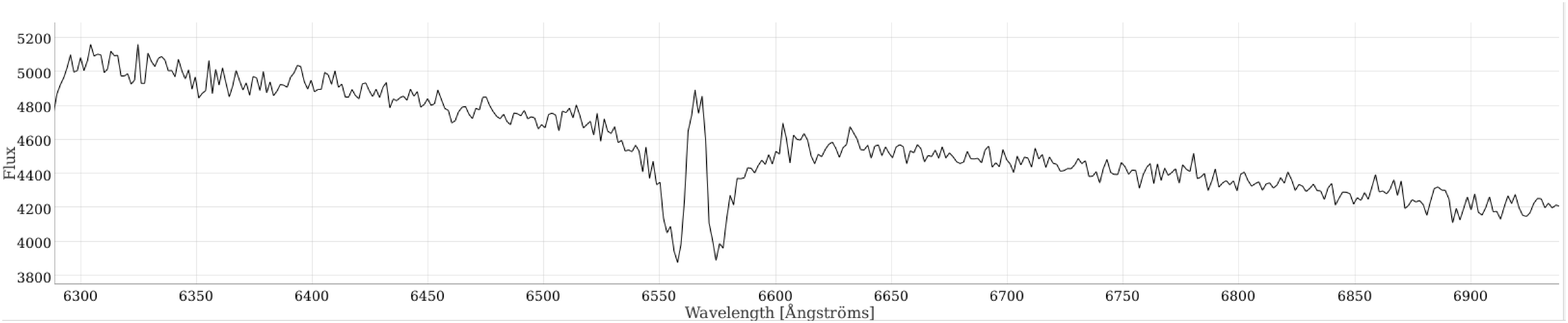} 
 \caption{Example of Be star identified in LAMOST DR1 by machine learning: \
Absorption line with weak central  emission peak}
   \label{absorption_emission}
\end{center}
\end{figure}

\begin{figure}[!ht]
\begin{center}
\includegraphics[width=\textwidth]{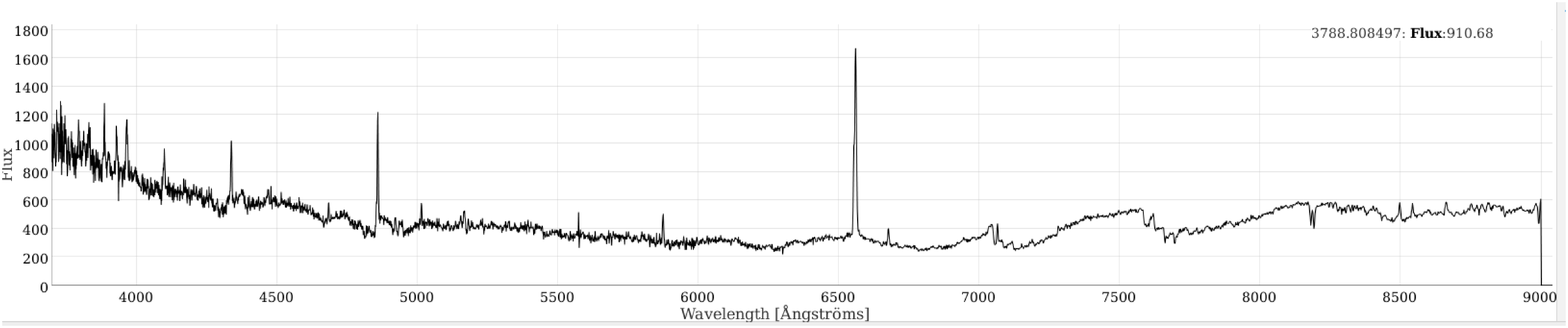} 
 \includegraphics[width=\textwidth]{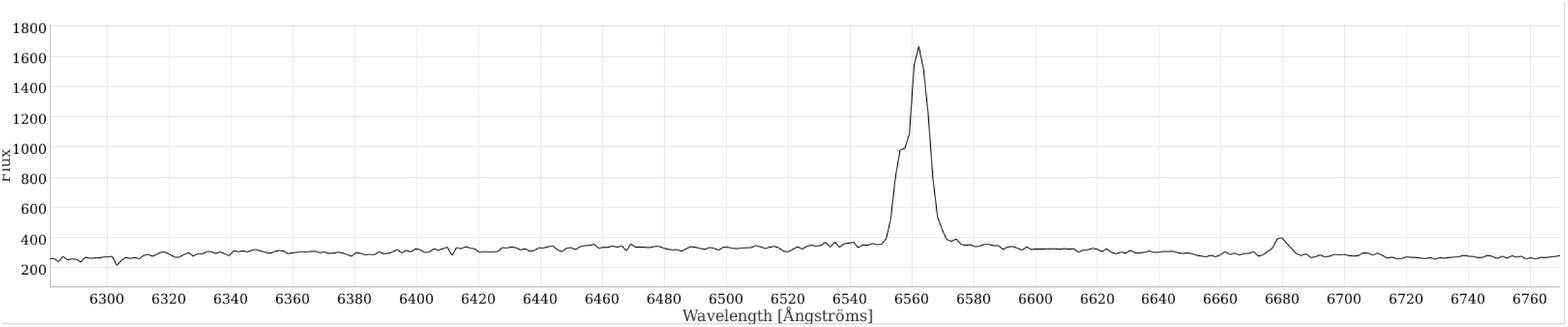} 
 \caption{LAMOST star with line profiles similar to Beta Lyrae}
   \label{betlyr_like}
\end{center}
\end{figure}

\begin{figure}[!ht]
\begin{center} \hspace*{-0.08\textwidth}
 \includegraphics[width=1.12\textwidth,height=0.4\textwidth]{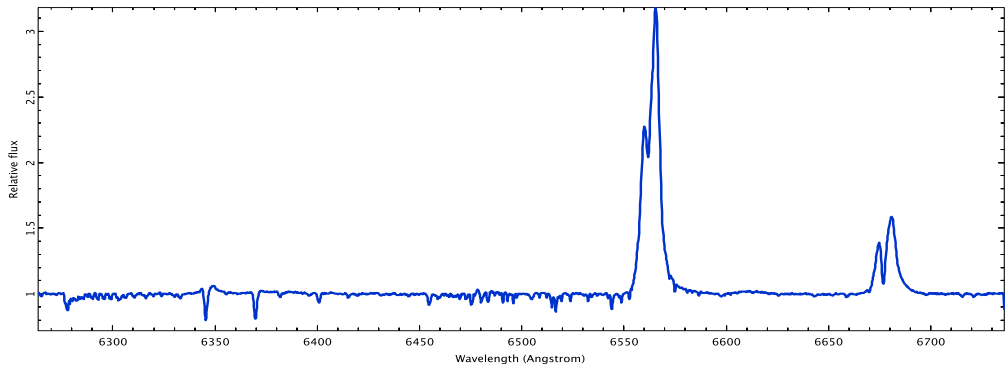} 
 \caption{Beta Lyrae from Ond\v{r}ejov CCD700 archive}
   \label{betlyr}
\end{center}
\end{figure}

The result of running the semi-supervised machine learning on a set of feature
vectors is a confidence score of each --- the number expressing the weight of
the label assigned by the algorithm to the unlabelled data. By thresholding the
highest confidences we get the most probable candidates of our target class (in
our case the emission line spectra).
Our experiments run many times of subsamples of LAMOST DR1 gave us a short list
of interesting spectra  which were previewed to eliminate artifacts. The final
list contains tens of very interesting candidates of emission line stars
deserving further investigation. 
On Fig.~\ref{double_emission} and Fig.~\ref{absorption_emission} are given examples of two different profiles of Be stars found by semi-supervised training (the whole spectrum and zoomed one centered around H$_\alpha$ line).
The Fig.~\ref{betlyr_like} shows another interesting object with emission both in Hydrogen and Helium 6678\AA{} lines. This configuration of emission  line profiles  closely resembles the well-known Beta Lyrae star. For comparison the spectrum of Beta Lyrae from Ond\v{r}ejov CCD700 archive is given on Fig.~\ref{betlyr}.
\section{Conclusions} 
Big spectral archives such as LAMOST DR1 are good source of data suitable for
machine learning of interesting objects according to their characteristic
spectral line shape.  Examples of objects with emission lines found by our
method confirm, that  the whole idea of machine learning applied to spectral line
profiles is viable and the methods described above are able to identify object
of given spectral properties.  The identified candidates with emission profile
need  further detailed examination as they may hide interesting  scientific
objects. The semi-supervised  methods may benefit considerably from massive
parallelisation using Spark on Hadoop cluster. 
\acknowledgements We are indebted to the support of grant LD-15113 of Ministry of
Education, Youth and Sports of the Czech Republic.  This research is based on
spectra from Ond\v{r}ejov 2m Perek telescope and public LAMOST DR1 survey.
Access to IT facilities of the National Grid Infrastructure MetaCentrum,
provided under the programme "Projects of Large Research, Development, and
Innovations Infrastructures" (CESNET LM2015042), is greatly appreciated. P.S.
also acknowledges the travel support of COST Action TD-1403 and of IAU office.
\vspace*{-3ex}


\begin{thebibliography}{}
%
\bibitem[Chapelle et al. (2006)]{chapelle2006semi}
  {Chapelle, O., Sch{\"o}lkopf, B., \&  Zien, A.} 2006,
  \textit{Semi-supervised learning},
  {MIT press, Cambridge, Massachusetts}
%
\bibitem[Cui et. al. 2012]{Cui}
{Cui, X.Q. et al} 2012, \textit{Research in Astronomy and Astrophysics}, 12, 1197
%
\bibitem[Luo et al. 2015]{Luo}
{Luo, A.L. et al.} 2015, \textit{Research in Astronomy and Astrophysics}, 15, 1095
%
\bibitem[(Nandrekar-Heinis et al. 2014)]{TAP} Nandrekar-Heinis, D., Michel, L., Louys, M., \& Bonnarel, F.\ 2014, {\textit Astronomy and Computing}, 7, 37 
%
\bibitem[Pali\v{c}ka (2016)]{Palicka_MT}
{Pali\v{c}ka, A.} 2016, \textit{Master Thesis}, 
Czech Technical University in Prague, Faculty of IT
%
\bibitem[({{Porter} \& {Rivinius} 2003})]{2003PASP..115.1153P}
{Porter}, J.~M., \& {Rivinius}, T. 2003, \textit{PASP}, 115, 1153
%
\bibitem[Shakurova (2016)]{Shakurova_MT}
{Shakurova, K.} 2016, \textit{Master Thesis}, 
Czech Technical University in Prague, Faculty of IT
%
\bibitem[({Silaj} et~al. 2010)]{2010ApJS..187..228S}
{Silaj}, J., {Jones}, C.~E., {Tycner}, C., {Sigut}, T.~A.~A., \& {Smith}, A.~D.
  2010, \textit{ApJS}, 187, 228
%
\bibitem[(Tody et~al. 2012)]{SSAP}
{Tody, D. et~al.} 2012, \textit{IVOA Recommendation: Simple Spectral Access Protocol Version 1.1}, ArXiv:1203.5725
%
\bibitem[{({Zickgraf} 2003)}]{2003A&A...408..257Z}
{Zickgraf}, F.-J. 2003, \textit{A\&A}, 408, 257
\end{thebibliography}
\end{document}